\documentclass[conference]{IEEEtran}
\IEEEoverridecommandlockouts
\usepackage{cite}
\usepackage{booktabs}   
\usepackage{subcaption} 
\usepackage[normalem]{ulem}
\usepackage{pifont}
\usepackage{amsmath}

\usepackage{amssymb}
\usepackage{amsfonts}
\usepackage{algorithm}
\usepackage{algpseudocode}
\usepackage{url}
\usepackage{hyperref}
\usepackage{comment}
\usepackage{color}
\usepackage{balance}
\usepackage{listings}
\usepackage{multirow}
\usepackage{enumitem}
\usepackage{mathtools}
\usepackage{graphicx}
\usepackage{pbox}
\usepackage{makecell}
\usepackage{rotating}
\usepackage{adjustbox}
\usepackage{amsthm}
\usepackage{mathtools}
\usepackage{xcolor}
\usepackage{textcomp}
\usepackage{soul}

\usepackage[compatibility=false,font=small,labelfont=bf]{caption}

\def\BibTeX{{\rm B\kern-.05em{\sc i\kern-.025em b}\kern-.08em
		T\kern-.1667em\lower.7ex\hbox{E}\kern-.125emX}}

\newcommand{\tool}[0]{\mbox{\textsc{EasyView}}}

\begin{document}

\title{\tool{}: Bringing Performance Profiles into Integrated Development Environments}

    
    
\author{
\IEEEauthorblockN{Qidong Zhao}
\IEEEauthorblockA{
\textit{North Carolina State University} \\
Raleigh, USA \\
qzhao24@ncsu.edu}
\and
\IEEEauthorblockN{Milind Chabbi}
\IEEEauthorblockA{
\textit{Scalable Machines Research} \\
San Francisco, USA \\
milind@scalablemachines.org}
\and
\IEEEauthorblockN{Xu Liu}
\IEEEauthorblockA{
\textit{North Carolina State University} \\
Raleigh, USA \\
xliu88@ncsu.edu}
}

\maketitle

\begin{abstract}
Dynamic program performance analysis (also known as profiling) is well-known for its powerful capabilities of identifying performance inefficiencies in software packages. Although a large number of profiling techniques are developed in academia and industry, very few of them are widely used by software developers in their regular software developing activities. There are three major reasons. First, the profiling tools (also known as profilers) are disjoint from the coding environments such as IDEs and editors; frequently switching focus between them significantly complicates the entire cycle of software development. Second, mastering various tools to interpret their analysis results requires substantial efforts; even worse, many tools have their own design of graphical user interfaces (GUI) for data presentation, which steepens the learning curves. Third, most existing profilers expose few interfaces to support user-defined analysis, which makes the tools less customizable to fulfill diverse user demands.

We develop \tool{}, a general solution to integrate the interpretation and visualization of various profiling results in the coding environments, which bridges software developers closer with profilers during the code development cycle. The novelty of \tool{} lies in its significant improvement on the usability of profilers. \tool{} not only provides deep insights to support intuitive analysis and optimization in a simple interface, but also enhances user experiences in using the profilers effectively and efficiently in the IDEs. 
Our evaluation shows that  \tool{} is able to support various profilers for different languages and provide unique insights into performance inefficiencies in different domains. Our user studies show that \tool{} can largely improve the usability of profilers in software development cycles via facilitating performance debugging efforts.

\end{abstract}

\begin{IEEEkeywords}
Profiling, Software optimization, Performance measurement, Visualization, Tools.
\end{IEEEkeywords}

\section{Introduction}
\label{sec:introduction}

Production software packages have become increasingly complex. 
They are comprised of a large amount of source code of different languages, sophisticated control and data flows, a hierarchy of component libraries, and growing levels of abstractions. 
This complexity often introduces inefficiencies across the software stacks, leading to resource wastage, performance degradation, and energy dissipation~\cite{Molyneaux:2009:AAP:1550832, Bryant:2010:CSP:1841497}.
The provenance of these inefficiencies can be many: rigid abstraction boundaries, missed opportunities to optimize common cases, suboptimal algorithm design, inappropriate data structure selection, poor compiler code generation, and problematic software-hardware interactions. Program analysis plays an important role in understanding performance inefficiencies and guiding code optimization.

There are two kinds of program performance analysis: static and dynamic. Static analysis typically leverages compiler infrastructures to study program source code, byte code, or binary code. 
Static analysis is adept at exploring performance inefficiencies via  techniques such as common subexpression elimination~\cite{deitz2001eliminating}, value numbering~\cite{gvn},  constant propagation~\cite{Wegman:1991:CPC:103135.103136}, among others. 
Orthogonal to static analysis is the dynamic program analysis (aka profiling) that identifies program inefficiencies at runtime.
Performance analysis tools (aka profilers) such as HPCToolkit~\cite{adhianto2010hpctoolkit}, VTune~\cite{vtune}, gprof~\cite{Graham-etal:1982:PLDI-gprof}, pprof~\cite{pprof}, OProfile~\cite{Levon:OProfile}, perf~\cite{perf}, and many others monitor code execution to identify hot code regions, idle CPU cycles,  arithmetic intensity, and cache misses, among others. 
Static analysis and dynamic analysis complement each other and are often used together for deep performance insights. 


While program performance analysis has been proven useful, having it widely adopted by software developers and continuously used in the regular software development cycles is difficult. The main reasons include (1) mastering various program analysis tools requires steep learning curves, especially for tools of different features; (2) customizing the analysis to fulfill the diverse needs is difficult; 
and (3) frequently switching focus between analysis tools and coding environment can significantly distract developers and complicate development cycles~\cite{10.1109/TSE.2006.116,interrupt}, especially when developers are immersed in mission-critical tasks~\cite{csikszentmihalyi2014flow,10.5555/951952.952340}. 

To address these challenges, many research efforts, such as MagpieBridge~\cite{magpiebridge}, IBM AppScan~\cite{xforce}, Xanitizer~\cite{whitesource},  aim to integrate program analysis into  interactive development environments (IDEs)  or editors. Most existing approaches focus on only the static analysis because the static analysis is usually based on compilers, which hide most of the language details with compiler front ends and intermediate representations. Moreover, language server protocol~\cite{lsp} is proposed to easily integrate static analysis into IDEs and editors. 

In contrast, there is no systematic solution for enhancing the usability of profilers in the development cycles. Today, profilers~\cite{perf,pprof,async-profiler,adhianto2010hpctoolkit,tau,dynatrace} are usually implemented as \emph{standalone} tools with their own data collection, analysis, and visualization. These profilers target different programming languages (e.g., C/C++, Java, Go), different application domains (e.g., cloud, high performance computing, mobile devices), different performance inefficiencies (e.g., hotspots, insufficient parallelism, memory bottlenecks), and different insights (e.g., various metrics, data/control flows, profiling, and tracing). However, users face to steep learning curves to master these profilers.
For improvement, some profilers are integrated into popular IDEs, such as Visual Studio~\cite{vs}, JetBrains products~\cite{jetbrain} (e.g., IntelliJ, Goland, CLion), and Eclipse~\cite{eclipse},  for in-situ analysis and visualization~\cite{ICPC-2013-BeckMDR}. However, existing IDE-based solutions lack in three aspects to prevent them from high usability. (1) Existing solutions limit IDEs to support only few profilers, so users may not be able to use their desired profilers. (2) Existing solutions do not take advantage of profiler-IDE integration; one does not have flexibility to extend and customize built-in profilers to improve the use experiences. (3) Existing approaches do not work efficiently to handle large profiles, so users may suffer from high latency in opening and exploring profiles.

To fill this gap, we develop \tool{}, which aims to improve user experiences of profiling tools by bringing profilers closer to software engineers for the regular use in development cycles.
\tool{} makes the following contributions.
\begin{itemize}[leftmargin=*]
\item \tool{} employs a generic solution. It unifies common features of mainstream profilers into a generic representation. As a result,  \tool{} supports profiles from a  wide range of analysis and visualization to fulfill different needs from software engineers.

\item \tool{} obtains deep insights. \tool{} supports customizable interfaces to produce unique views to advance state-of-the-art analysis and visualization of performance data, which enables the most flexibility for users to enjoy deep performance insights.

\item \tool{} enhances user experiences. \tool{} tightly integrates the analysis with Visual Studio Code~\cite{vscode}, supporting efficient profile analysis and visualization. Users can enjoy low response time and  smooth interactions in exploring the profiling data together with source codes.

\end{itemize}

We evaluate \tool{} with both case and user studies. The case studies show that \tool{} requires minimum coding efforts to support different profilers and enables new performance analysis that provides unique insights for application optimization in various domains. Moreover, we show \tool{} outperforms prior approaches in efficient data processing and visualization. The user studies with experienced software engineers show that \tool{}, with tight integration into the coding environment, can largely facilitate the code analysis in their daily development activities. We also show the effectiveness of \tool{} by creating control groups to evaluate the user experiences with \tool{} and the existing standalone or IDE-based solutions. Experimental results show that \tool{} can significantly improve usability of profilers for software engineers.


\begin{figure*}[t]
  \centering
  \includegraphics[width=.75\textwidth]{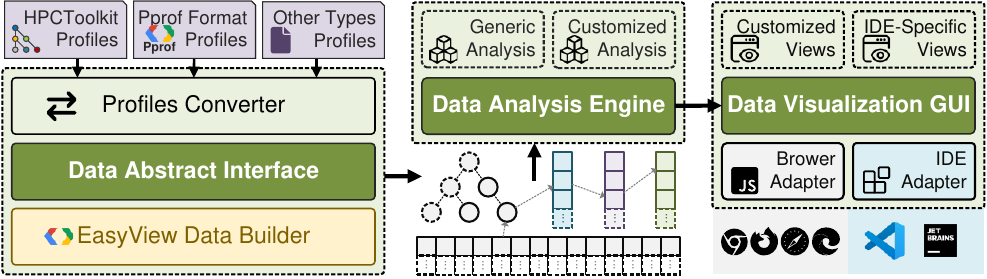}
  \centering
  \caption{Overview of \tool{}. \tool{} consists of three components: profile abstraction, analysis, and visualization. } 
  \label{fig:overview}
\end{figure*}

\section{Background and Related Work}
\label{relate}

Data analysis and visualization are important components of performance engineering tools. There are typically three mechanisms to present the data to users: text format, dedicated graphical user interface (GUI), and plugins in IDEs/editors. The text format is widely used in research tools~\cite{loadspy,jxperf,valgrind} developed in academia, which is easy to implement and evaluate. Some tools, such as perf~\cite{perf}, pprof~\cite{pprof}, and Scalene~\cite{berger2020scalene}, improve the text format via the Markdown format~\cite{markdown}, which can be visualized in a shell terminal or a web browser with being translated to HTML. 
However, the text, markdown, and HTML formats lack flexibility in analyzing the performance data, such as 
deriving new metrics, customizing views, and presenting large codebases in a scalable way. 

Instead, many research and commercial tools utilize dedicated GUI to visualize their analysis results for more flexibility. For example, tools such as HPCToolkit~\cite{adhianto2010hpctoolkit}, VTune~\cite{vtune}, hotspot~\cite{hotspot}, TAU~\cite{tau}, and Oprofile~\cite{Levon:OProfile} have their own GUIs written in Java or Qt.  Google Cloud Profile~\cite{googleprofiler}, Perfetto~\cite{perffeto}, SpeedScope~\cite{speedscope}, Pyroscope~\cite{Pyroscope}, gProfiler~\cite{gProfiler}, pprof~\cite{pprof}, and FlameScope~\cite{Flamescope} visualize the performance data interactively in web browsers.  However, these dedicated data visualizations suffer from four weaknesses, which impede them from wide adoption. First, installing these visualizers and learning their usage incurs extra overhead for programmers. Second, according to prior studies~\cite{10.1109/TSE.2006.116,interrupt,csikszentmihalyi2014flow,10.5555/951952.952340}, the stand-alone GUIs require users to switch between the tool and code editors/IDEs, which significantly delays the development process.
Third, these dedicated GUIs lack interoperability across profiles produced by different tools and  are not designed to support customizable analysis. Fourth, some web-based approaches such as Pyroscope (i.e., flamegraph.com) require uploading the profiles to their servers, which raises some security and privacy concerns.

To address these limitations, \tool{} adopts a plug-in approach in IDEs and code editors. Compared to other approaches, it provides in-situ visualization of performance in the same IDE where developers write, test, and debug code. In the rest of this section, we first introduce VSCode on which \tool{} is built as a plug-in and then compare to existing approaches that integrate profiling into code editors/IDEs. 

\subsection{VSCode}

Microsoft’s Visual Studio Code (VSCode)~\cite{vscode} is a text editor with powerful IDE-like features. VSCode supports a wide range of programming languages and is highly customizable with various extensions, which are designed for both beginners and advanced programmers. VSCode has been a popular editor in the community.
VSCode enjoys the following features: (1) VSCode is free to download and simple to install on most operating systems (e.g., Windows, macOS, Linux); (2) VSCode is easy to customize and supports many useful plug-in extensions. \tool{} leverages these features of VSCode to enjoy the general, extensible, and applicable analysis. Similar to VSCode, IDEs/editors such as JetBrains products, Atom, Eclipse provide similar plug-in capabilities.

\subsection{Related Work on Program Analysis in IDEs}

While many performance analysis and visualization techniques are developed, we only review the most related approaches that tightly integrate the program analysis into IDEs or code editors. 
Some existing approaches integrate static analysis into IDEs and code editors to improve usability.
For example, the widely used Language Server Protocol (LSP)~\cite{lsp} defines the protocol between an IDE and a language server that provides language features like auto-completion, goto definition, and finding all references. 
With the support of LSP, tools such as  MagpieBridge~\cite{magpiebridge} integrate static analyses into IDEs. However, LSP is not designed for profilers; it does not handle various formats and internals of program profiles.

For profiling, almost all the mainstream editors/IDEs have integrated profilers.
JetBrains integrates Async-profiler for its IntelliJ IDEA~\cite{intellij}, Perf and DTrace for CLion~\cite{clion}, and PProf for Goland~\cite{goland}. 
For VSCode, a variety of profilers, such as VTune~\cite{vtune-vscode}, PProf~\cite{pprof-vscode}, and Austin~\cite{austin-vscode} implement their visualization interfaces as extensions for tight integration.  Moreover, existing approaches~\cite{performancehat,10.1145/3173574.3174106,10.1145/2556288.2557409,ICPC-2013-BeckMDR} augment the source code views in IDEs with integrating profiling information. 
These existing approaches are limited to a few individual profilers, with no general and interoperational solutions.
Moreover, these approaches simply show the traditional stand-alone views in IDEs, without fully taking advantages of integrations such as various annotations on codes and actions on profiles.
In contrast, \tool{} provides a systematic solution that goes beyond the simple improvement over some individual profilers. \tool{} supports general, extensible, insightful, and efficient analysis across multiple profilers, which no existing approaches can easily obtain. 

\section{\tool{} Overview and Scope}
\label{overview}
Figure~\ref{fig:overview} overviews \tool{}, which consists of a data abstraction interface, a data analysis engine, and a data visualization GUI. \tool{} follows four design principles.
\begin{itemize}[leftmargin=*]
\item General. \tool{} is not designed for supporting specific profilers. Instead, it aims to provide a general solution that widely supports a broad range of profilers. \tool{} designs general data representation, supports general analysis, and presents profiles in general views.
\item Customizable. \tool{} is designed to be highly flexible. It supports customized or personalized analysis and visualization to better fulfill the diverse demands in different domains. This enables \tool{} to integrate users' knowledge or various data mining and machine learning techniques to maximize insights.
\item Applicable. \tool{} is implemented as an IDE/editor's plug-in with web front techniques (e.g., TypeScript, JavaScript, Web Assembly), which is portable on different platforms such as Linux, Windows, and macOS. Furthermore, \tool{} analyzes and visualizes data locally without uploading data to a remote server, which minimizes the security and privacy concerns.
\item Efficient. \tool{} minimizes the overhead of processing profiling data, so users have smooth experiences (i.e., low response time) when handling large profiles.
\end{itemize}

\textit{Scope:} \tool{} focuses on analyzing and visualizing profile data but not on profile data collection. 
The goal of \tool{} is not to replace existing profilers; instead, \tool{} aims to improve the usability of most (if not all) existing profilers by bringing them closer to the development environment familiar to the software --- IDEs. IDEs can also easily support various performance tools via \tool{}, not limited to their built-in profilers only.
We elaborate on different components of \tool{} in the following sections.


\section{\tool{}'s Data Abstraction Interface}
\label{format}

With this component, \tool{} abstracts the performance data collected from different tools into a unified representation. We first define a generic data representation that unifies common features of mainstream profilers and then develop a set of APIs to bind the generic data representation with existing and emerging profilers. 

\subsection{Generic Data Representation}
\label{representation}
To design the generic data representation, we study more than 50  profilers, which are both mainstream and state-of-the-art. 
These profilers cover different domains, such as high performance computing (e.g., HPCToolkit~\cite{adhianto2010hpctoolkit}, TAU~\cite{tau}, ScoreP~\cite{scorep}, Caliper~\cite{caliper}),  system and microarchitecture (e.g., Intel VTune~\cite{vtune}, ARM MAP~\cite{armmap}, AMD uProf~\cite{uprof}, NVIDIA Nsight Compute~\cite{nsight}, Linux Perf~\cite{perf}, and Oprofile~\cite{Levon:OProfile}), high-level languages, e.g., Python, Java, Go (
Async-Profiler~\cite{async-profiler}, JXPerf~\cite{jxperf}, PProf~\cite{pprof}, Scalene~\cite{berger2020scalene}), and fine-grained analysis (e.g., Valgrind~\cite{valgrind}, CCTLib~\cite{cctlib}, DrCCTProf~\cite{drcctprof}, Pin~\cite{Luk:2005:PBC:1065010.1065034}, NVBit~\cite{nvbit}).
We identify the following common features owned by most profilers.
\begin{itemize}[leftmargin=*]
\item \emph{Profiling contexts:} Profilers typically analyze and report the performance insights of code regions at the granularity of the entire program, functions, loops, basic blocks, or individual instructions, which are known as profiling contexts.
\item \emph{Metrics:} Profilers always provide one or more metrics, such as time, cycles, memory consumption, cache misses, lock contention, and many others. These metrics are associated with profiling contexts and used to rank performance issues.
\item \emph{Call paths:} Many profilers provide call paths (also known as calling contexts, backtraces) to give additional analysis insights. A calling context consists of a series of frames on the call stack at any given profiling context. 
\item \emph{Code mapping:} In order to provide actionable optimization guidance, profilers map the analysis results to programs at the binary or source code levels. The mapping requires instruction pointers, load modules and offsets, source code files and their paths, and line numbers.
\end{itemize}

\begin{figure}[t]
  \begin{center}
  \includegraphics[width=0.3\textwidth]{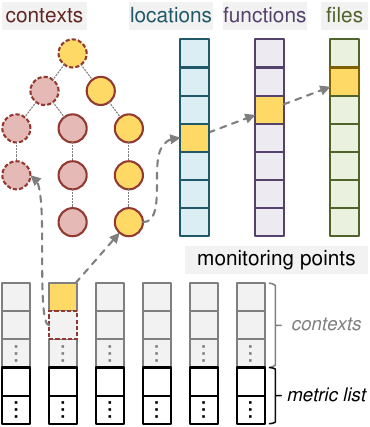}
  \end{center}
  \caption{The profile representation used by \tool{} is designed as a tree structure to maintain call path profiles. The representation is expressed in a Protocol Buffer schema (hide for review).} 
  \label{fig:format}
\end{figure}

\tool{} encodes these features into a generic data representation, as shown in Figure~\ref{fig:format}. All the monitoring points are organized into a compact calling context tree (CCT) by merging the common prefixes of their call paths,  which minimizes the storage in both memory and disk for the profiles. For each monitoring point, \tool{} maintains two pieces of information: {\tt context} and {\tt metric list}; {\tt context} points to the corresponding CCT node with the source code attribution (line location, function, and file), and {\tt metric list} points to the list of metrics associated with this monitoring point. \tool{}'s data representation, expressed in a Protocol Buffer schema~\cite{proto} supports all the aforementioned common features. 

Besides expressing these common features, \tool{}'s representation also supports advanced features owned by different tools. First, the profiling contexts represent not only the traditional code regions but also data objects, such as heap objects in their allocation call paths and static objects with their names in the symbol tables, so it can handle many memory profilers with data-centric analysis, such as Perf, ScaAnalyzer~\cite{Liu:2015:STI:2807591.2807648}, DrCCTProf~\cite{drcctprof}, Cheetah~\cite{Liu:2016:CDF:2854038.2854039}, MemProf~\cite{memprof}. Second, \tool{}'s representation is able to associate multiple metrics with a monitoring point. Moreover, \tool{} can associate multiple contexts and monitoring points to a single metric. 
This capability is particularly useful for profilers that identify performance inefficiencies involving multiple contexts, for example, data reuse~\cite{Marin-sweep3d} with use and reuse contexts, computation redundancy~\cite{loadspy,redspy,witch} with redundant and killing contexts, data races with two memory access contexts, and false sharing~\cite{feather} with memory accesses in two contexts ping-ponging. 
We show how these features support powerful analysis in Section~\ref{hpc}.

\subsection{Data Binding}

\tool{} binds its representation with the performance data produced by various tools. However, developing a general solution to fulfill most existing profilers is challenging. On the one hand, tools usually have their own data formats based on binary (e.g., PProf, Perf, HPCToolkit) or JSON (e.g., Chrome Profiler). On the other hand, different languages are used to develop profilers; for example, Perf uses C and PProf uses Go. These diverse formats and languages complicate the applicability of \tool{}'s representation.

To address these challenges, 
\tool{} employs a data builder, which (1) derives simple high-level APIs from the functions generated by Protocol Buffers and (2)  binds high-level APIs with different languages, such as C, C++, Python, Go, to name a few. Existing profilers use the data builder in two ways. On the one hand, profilers can directly leverage the data builder to output data in \tool{}'s representation. The examples include some open-source tools, such as DrCCTProf~\cite{drcctprof} and JXPerf~\cite{jxperf}. On the other hand, \tool{} provides a format converter atop the data builder, which translates the outputs of existing profilers to \tool{}'s representation. This mechanism avoids major changes to existing profilers, which broadens the support for a wide range of existing profilers. Currently, \tool{}'s format converter supports PProf~\cite{pprof}, Perf~\cite{perf}, Cloud Profiler~\cite{googleprofiler}, Scalene~\cite{berger2020scalene}, Chrome profiler~\cite{chromeprofiler}, HPCToolkit~\cite{adhianto2010hpctoolkit}, TAU~\cite{tau}, and pyinstrument~\cite{pyinstrument}; the list is expanding rapidly. We show that using \tool{} data builder directly or for conversion requires minimum engineering efforts in Section~\ref{eval1}.

\section{\tool{}'s Data Analysis Engine}
\label{analysis}

In this component, \tool{} analyzes the performance data in its representation. To enable broad applicability, the analysis engine is implemented with pure web front-end techniques (i.e., JavaScript, TypeScript), so we can easily deploy the analysis together with the GUI in VSCode, with no need for additional software installation. In the rest of this section, we describe \tool{}'s engine for both general and customized analyses. 

\subsection{General Data Analysis}
\label{general}

As shown in Figure~\ref{fig:format}, \tool{} represents the performance data into a tree data structure, so \tool{} operates the tree for several common analyses, which is backward compatible with existing profilers.

\paragraph{Tree traversal}
\tool{} supports basic tree traversal operations, which iterate all tree nodes in different orders. Associated with the tree traversal, \tool{} performs corresponding analysis, such as computing inclusive/exclusive metrics, collapsing deep and recursive call paths, and pruning insignificant tree nodes. 

\paragraph{Tree transformation}
\tool{} transforms the tree into top-down, bottom-up, and flat shapes for more insights. The top-down tree, rooted at the entry function (e.g., {\tt main} or {\tt thread\_main}) with callees as children, shows how the metric distribution along the call paths. The bottom-up tree, which reverts the top-down tree, shows the hot functions called in various call paths. The flat tree, regardless of call paths, shows the hot load modules (e.g., shared libraries), files, and functions. \tool{} visualizes all the three shapes of the tree to enable powerful analysis. 

\paragraph{Operations across multiple profiles}
\tool{} is able to analyze multiple profiles via managing multiple tree structures, which is particularly useful for tools that produce separate profiles for different threads, processes, or executions.
\tool{} supports two basic operations: {\em aggregation} and {\em differentiation}. The aggregation operation merges the profiles by constructing a unified tree and deriving statistical metrics associated with each node in the unified tree. 
\tool{} maintains the metrics from all the profiles and predefines some operators to derive metrics from them, such as {\tt sum}, {\tt min}, {\tt max}, and {\tt mean} across different profiles. The aggregation enables \tool{} to correlate multiple profiles and show a compact view.

The differentiation operation quantifies the difference between two profiles collected in two different executions, which provides unique insights~\cite{Liu:2015:STI:2807591.2807648}, such as scaling losses and resource contention. The differentiation operation is similar to the aggregation operation. By default, two nodes are differentiable if all the parents (ancestors) are differentiable.
Unlike existing approaches~\cite{Pyroscope}, \tool{} shows detailed differences in both tree nodes and metrics in all the top-down, bottom-up, and flat trees. 

\subsection{Customized Data Analysis}

To allow extensions to the data analysis, \tool{} exposes a programmable interface, as a programming pane in the GUI, for users to customize the analysis. Based on Python-WASM~\cite{python-wasm}, users write Python codes to customize the profile data in the pane, which can be translated to Web Assembly for direct execution in \tool{}. It is worth noting that \tool{} does not require any additional software installation or server setup to enable the customizable analysis. To minimize the manual efforts on devising the customizable analysis, \tool{} triggers user-defined callback functions in the tree operations defined in Section~\ref{general}. There are mainly two types of callback functions.
\begin{itemize}[leftmargin=*]
\item Callbacks at node visit. Upon visiting each node during tree traversal or transformation, \tool{} triggers a callback for users to define how to process the current node. For example, users can decide to merge two nodes if they are mapped to the same source code line. Moreover, users can elide any nodes in the tree that are not of interest.
 
\item Callbacks at metric computation. \tool{} triggers a callback to allow users to define any formula to derive new metrics. For example, users can compute cycles per instructions, cache misses per thousand instructions, and many others via specifying the corresponding formulae. Moreover, users can use division instead of subtraction to derive differential metrics, which is used to measure memory scaling~\cite{Liu:2015:STI:2807591.2807648}.
\end{itemize}

\tool{} exposes traversals over the internal tree, so users can manipulate the profile via accessing any nodes and metrics. One can easily integrate more data mining or machine learning techniques to analyze the profiles.

\subsection{Optimization for Efficiencies}
For efficient data analytics, \tool{} adopts WebAssembly and WebGL extensively with thorough optimization. \tool{} manages the memory manually to avoid frequent invocation of garbage collectors. Moreover, \tool{} avoids unnecessary data movement and computation in transforming or traversing the trees. In Section~\ref{efficiency}, we show \tool{} significantly outperforms existing approaches in response time with large profiles.

\section{\tool{}'s Visualization Interfaces}
\label{visualization}

\tool{} visualizes the profiles produced by its analysis engine. \tool{} supports generic views integrated with IDE-specific features to present the profiles. Like LSP~\cite{lsp}, \tool{} defines a set of activities that correlate views with source code in any IDEs.



\subsection{Customizable Views}
\label{cusview}
\begin{figure*}[t]
  \centering
  \includegraphics[width=.9\textwidth]{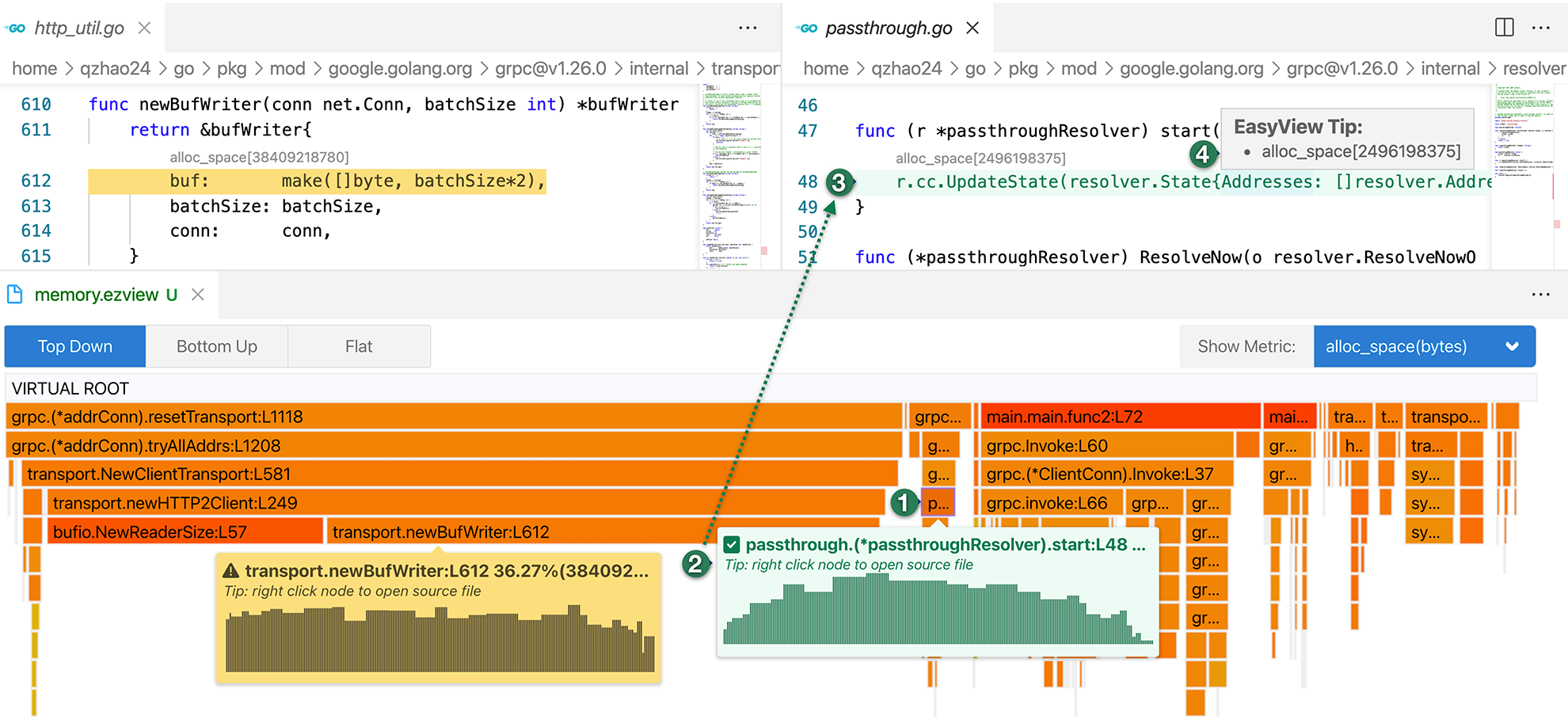}
  \centering
  \caption{\tool{} shows an aggregate memory profile for a gRPC application, collected by PProf. This snapshot overviews \tool{} that consists of flame graphs, source code annotations, and metric histogram. This profile identifies potential memory leaks. One can explore this profile in four steps. \textcircled{1} Select a frame in the flame graph. \textcircled{2} Click the frame to show the histogram. \textcircled{3} Right click the frame to link the source code in the top right pane. \textcircled{4} Place the cursor on the highlighted source code line to trigger the hover with detailed metrics.}
  \label{fig:grpc}
\end{figure*}

\paragraph{Generic views}
\sloppy
The default views of \tool{} are flame graphs~\cite{flamegraph}, which are the de facto view for performance data widely used by the community. Figure~\ref{fig:grpc} shows a typical flame graph produced by \tool{} based on the top-down tree produced by the analysis engine, which is called as a top-down view. The flame graph succinctly represents the tree structure maintained by \tool{}: the root represents the program entrance (i.e., process or thread start); the nodes underneath represent the call paths; the length of each node denotes the inclusive metric value. Besides the top-down flame graph, \tool{} also presents another two variants.

\begin{itemize}[leftmargin=*]
\item Bottom-up flame graph.
Based on the bottom-up tree, \tool{} constructs the bottom-up flame graph, which reverses the call path to have callees as parents and callers as children. \tool{} computes both the inclusive and exclusive metrics and shows them in two flame graphs. The bottom-up view is particularly useful to identify hot functions and understand where they are called from.

\item Flat flame graph.
Based on the flat tree, \tool{} elides all the call path information to provide a flat view. The flame graph is organized as follows in a hierarchy: the entire program, load modules (executable binary and shared libraries), files, and functions. Similar to the bottom-up flame graph, \tool{} also shows inclusive and exclusive metrics in two flat flame graphs. This flat view can highlight the hot shared libraries, files, and functions for optimization.
\end{itemize}

Moreover, all the flame graphs are searchable. One can highlight any function in the flame graph by searching the function name. While the design of these flame graphs is not new, \tool{} supports them for backward compatibility of existing approaches. Thus, \tool{} can attract users who are used to existing profilers.

\paragraph{Advanced views with customizable flame graphs}
Users can easily customize the flame graphs shown in \tool{} to better visualize the profiles. 
Currently, \tool{} supports three advanced flame graphs that (1) correlate multiple profiles, (2) aggregate multiple profiles, and (3) differentiate two profiles. Compared to existing approaches, these advanced views are novel in visualizing data in a more intuitive way.
First, \tool{}, with the support of the profile representation, can correlate contexts across multiple profiles. \tool{} presents this correlation in a flame graph variant. Figure~\ref{fig:lulesh-d-ru} shows an example, which correlates a context of a memory allocation in a program with all call paths in which the program accesses this memory. This visualization advances the solution used by state-of-the-art memory profilers of similar functionality~\cite{Liu:2015:STI:2807591.2807648,drcctprof}. Section~\ref{hpc} shows a use case of this advanced view.

Second, \tool{} visualizes the metric distribution from multiple profiles in an aggregate view. For any context in the aggregate profile, \tool{} attaches a histogram to show all the metrics of the same context from different profiles. This view is particularly useful to investigate the behavior across different threads/processes or across different runs. We show a case study in Section~\ref{cloud} with this advanced view.

\begin{figure}[t]
  \begin{center}
  \includegraphics[width=0.42\textwidth]{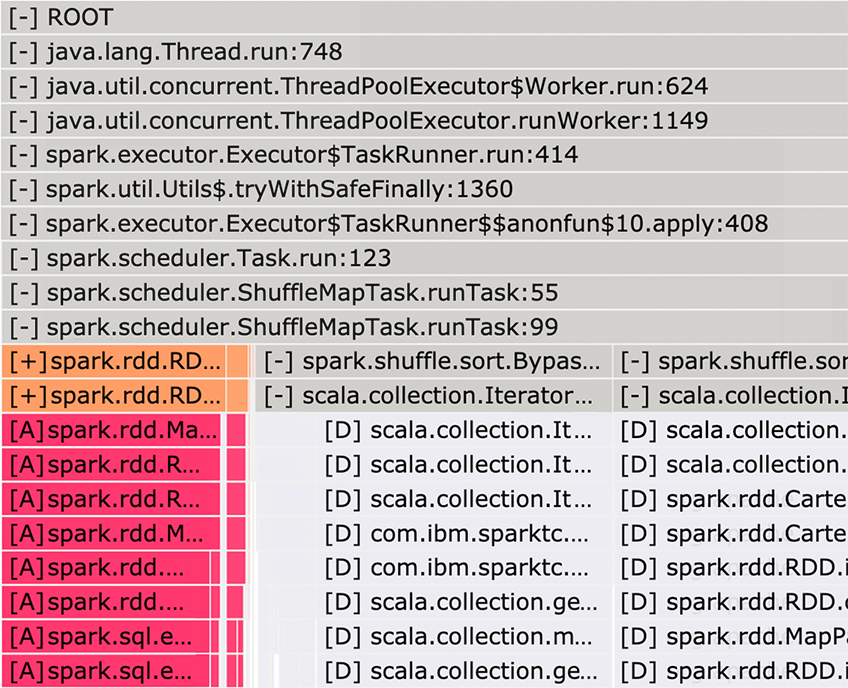}
  \end{center}
  \caption{\tool{}'s differential view on two profiles with a top-down flame graph. The prefixes in brackets denote the types of difference, and the flame graph quantifies the difference.} 
  \label{fig:diff}
\end{figure}

Third, the differential view is a special case of the aggregate view.
\tool{} compares two profiles in a differential flame graph, which is particularly useful to quantify the performance impact of the code or execution parameter changes. Compared to existing approaches~\cite{Pyroscope,flamegraph}, which only differentiate top-down flame graphs and use colors to provide a qualitative view, \tool{}'s differential flame graphs provide more insights into all the three types (i.e., top-down, bottom-up, and flat) of flame graphs and quantify the differences. As shown in Figure~\ref{fig:diff}, \tool{}'s differential flame graph shows four tags to compare profiles $\mathcal{P}_1$ and $\mathcal{P}_2$. {\tt [A]} means the newly added contexts in $\mathcal{P}_2$ but not existing in $\mathcal{P}_1$; {\tt [D]} means the newly deleted context in $\mathcal{P}_2$ but existing in $\mathcal{P}_1$. The prefixes {\tt [+]} and {\tt [-]} mean that the context exists in both $\mathcal{P}_1$ and $\mathcal{P}_2$; {\tt [+]} means the metric associated with the context is larger in $\mathcal{P}_2$ compared to $\mathcal{P}_1$, while {\tt [-]} means the metric associated with the context is smaller in $\mathcal{P}_2$ than in $\mathcal{P}_1$.  Figure~\ref{fig:diff} shows an example of differential profiles  collected by Async-Profiler~\cite{async-profiler} for Spark~\cite{spark} running with Spark-Bench~\cite{spark-bench}. $\mathcal{P}_1$ uses the RDD APIs~\cite{rdd} and $\mathcal{P}_2$ uses the SQL Dataset APIs~\cite{dataset}. 
From the figure, we can clearly see that SQL DataSet APIs outperform RDD APIs. The flame graph shows that the performance gains are from using an efficient SQL engine and bypassing costly data shuffle in RDD APIs.

\paragraph{Other views}
Besides flame graphs, \tool{} also supports the tree table view, which is another mainstream view in many mainstream profilers, such as  VTune, HPCToolkit, and TAU. \tool{}'s tree table view supports all the top-down, bottom-up, and flat views. Compared to flame graphs, the tree table view is less straightforward as it requires users to manually unfold any call paths; but it is particularly useful to visualize a profile with multiple metrics. In our user study, we compare the tree table views with the flame graphs and evaluate their effectiveness.

\subsection{IDE-enhanced Views}
Inspired by LSP, \tool{} gives the first efforts on defining a set of actions to annotate source code with profiling data shown in IDEs, which significantly improve user expriences. 

\paragraph{Mandatory actions}
Code link is the only mandatory action required by \tool{}, which links the profile with the source code. By clicking a block in the flame graph or a frame in the tree table, the IDE can open the corresponding source code file, jump to the line, and highlight it with a background color if the line mapping information is available in the profile.

\paragraph{Optional actions}
There are several actions that are unnecessary for \tool{} to integrate the profiles within the IDEs, but can facilitate the users to interpret the profiles.
\begin{itemize}[leftmargin=*]
\item Color semantics. \tool{} can define different colors upon different properties in the source code. \tool{}'s flame graphs can use different colors to represent profiles from different files or libraries and use different darkness to represent the availability of source line mapping.
\item Code lens. \tool{} can provide additional insights with code lens, which are annotations above (or below) the source code statements. \tool{} uses the code lens to show the assembly instructions associated with the source code statement (if the profile provides this information) and the metric values, as some profilers are designed for compiler developers to collect and maintain such assembly-level information.  
\item Hovers. \tool{} can generate any optimization tips and show them in hovers. Hovers can be associated with individual source lines and pop up when the mouse cursor is placed on the code. While \tool{} currently only shows all metric values associated with the selected line in the hovers, it opens an interface to record any advanced analysis results and show the optimization guidance with user-defined analysis.
\item Floating windows. \tool{} is able to open a floating window in the source code pane to summarize the entire profile. Unlike hovers, the floating windows provide the global summary of the entire program.
\end{itemize}

These actions are generally supported by mainstream IDEs. We implement \tool{} as an extension of Microsoft Visual Studio Code. \tool{} can be easily integrated into JetBrains products with its platform SDK~\cite{intellijsdk}.

\section{Evaluation}
\label{evaluation}

We evaluate \tool{} on the following fronts: (1) the efforts of using \tool{}'s profile representation (i.e., programmability), (2) the response time of \tool{} (i.e., efficiency), (3) the insights \tool{} can provide to optimize programs with mainstream profilers in multiple domains (i.e., effectiveness), and (4) the user experiences of using \tool{} (i.e., user studies).

\subsection{Programmability of \tool{}}
\label{eval1}
Since the design of \tool{} is to support a generic visualization for existing profilers, we evaluate the programming efforts of generating \tool{} data representation.
We quantify the lines of code needed for existing profilers to produce \tool{} data representation. There are three methods to adapt a profiler to support \tool{}:  (1) using \tool{}'s APIs to output the format used by \tool{} directly (e.g., DrCCTProf and JXPerf), (2) converting to \tool{}'s data representation (e.g., HPCToolkit, TAU, PyInstrument), or (3) using PProf data format, which is a subset of \tool{} representation in Protocol Buffer (e.g., perf, PProf, Cloud Profiler). We find that changing tools to directly output \tool{} format requires less than 20 lines of code (C++ for DrCCTProf and Python for JXPerf); converting to \tool{}'s format requires less than 200 lines of codes (e.g., C for Perf and Python for TAU/HPCToolkit) and most of them are used to parse the original profile formats. 

\begin{figure}[t]
  \begin{center}
  \includegraphics[width=0.45\textwidth]{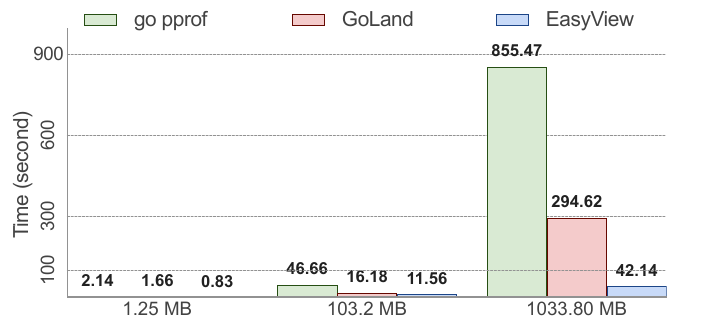}
  \end{center}
  \caption{Comparing response time among \tool{}, GoLand, and PProf. Lower is better.} 
  \label{fig:performance}
\end{figure}

\subsection{Efficiency of \tool{}}
\label{efficiency}
We measure the response time, which is defined as the end-to-end time of \tool{} to open a profile, including data processing (e.g., creating trees and computing metrics) and data visualization (e.g., rendering flame graphs). We glean real profiles collected by PProf data collector for industrial production software packages. The profile size ranges from $\sim$1MB to $\sim$1GB. To be fair, we have all the tools generate a top-down flame graph. Figure~\ref{fig:performance} shows the comparison among \tool{}, default PProf, GoLand of PProf plugin. We can see that \tool{} is much more efficient and others, especially for handling large profiles. 

\subsection{Effectiveness of \tool{}}
\label{optimization}
We evaluate the effectiveness of \tool{} by demonstrating how it can help obtain deep insights with the profilers in various domains.
We perform several case studies by two graduate students. These students know various profiling techniques but are not familiar with \tool{} or the applications under study.  
We show the case studies in both cloud and HPC domains, with the mainstream profilers and representative workloads. 
\emph{We show how \tool{}, with its unique capabilities, gives better optimization guidance. We also compare the usability of \tool{} and the profilers' default GUIs to show the advantages of \tool{}. }

\subsubsection{\tool{} in the Cloud Domain}
\label{cloud}

We study Go, one of the most popular program languages in cloud.
The profiler used in this study is PProf~\cite{pprof}, which is the de facto profiler for measuring Go programs.
We show that \tool{} can be easily integrated into the profiling workflow to facilitate the analysis efforts.
We study rpcx-benchmark~\cite{rpcx}, a popular test suite for gRPC~\cite{grpc} written in Go. This application uses the server-client model to simulate the high concurrency in practice. In this study, we show the profiling results in the clients. 
We follow the common usage of memory profiling in PProf to pinpoint potential memory leaks~\cite{memleaks}. We use PProf to periodically (i.e., every 0.1 second) capture a memory snapshot as a profile during the execution. Each snapshot shows the active memory usage in the allocation call paths. By analyzing the active memory consumption in each snapshot along the timeline, we can identify patterns that are due to potential memory leaks. PProf only identifies suspicious leaks, which require further investigations to confirm.  

\tool{} can analyze all the snapshots by aggregating them (with the technique described in Section~\ref{cusview}), as shown in Figure~\ref{fig:grpc}.  In this figure, the top two panes show the source code as well as the metrics in code lens and hovers supported by \tool{}; the bottom pane shows the profiles in flame graphs. We show the top-down flame graph, but one can also select bottom-up and flat flame graphs to show. One can right-click any part of the flame graph to associate the clicked frame with the source code. On the top right of the bottom pane shows the metric, which is the total allocated bytes in the profile. \tool{} can clearly highlight the hot memory allocations at {\tt bufio.NewResearSize} and {\tt transport.newBufWriter}, which are invoked when creating new HTTP clients (observed from the call path). When clicking the frame of {\tt transport.newBufWriter} in the flame graph, a hover popped up to show the histogram of active memory usage across different snapshots along the timeline. This histogram shows a pattern: the active memory in this call path is continuously high with no clear sign of reclamation. According to prior studies, this pattern raises a warning of potential memory leaks. Similarly, {\tt bufio.NewResearSize} also suffers from potential memory leaks. These memory leaks can be potentially caused by the clients not closing the connections in time. We have reported this finding to the application developers for further investigation. In contrast, the histogram in the right hover indicates no memory leaks in function {\tt passthrough} as the active memory usage is diminishing at the end of the program execution. 

\textit{Comparing to off-the-shelf PProf:} We obtain the similar analysis with the built-in PProf analysis. While PProf can collect and visualize profiles, it does not support automatic analysis across multiple profiles. Thus, one needs to write scripts to extract the data and manually plot the histogram of active memory usage for a given allocation context. In contrast, \tool{} significantly facilitates these efforts with its powerful analysis and visualization interfaces. 

\subsubsection{\tool{} in the HPC Domain}
\label{hpc}

\begin{figure}[t]
  \begin{center}
  \includegraphics[width=0.42\textwidth]{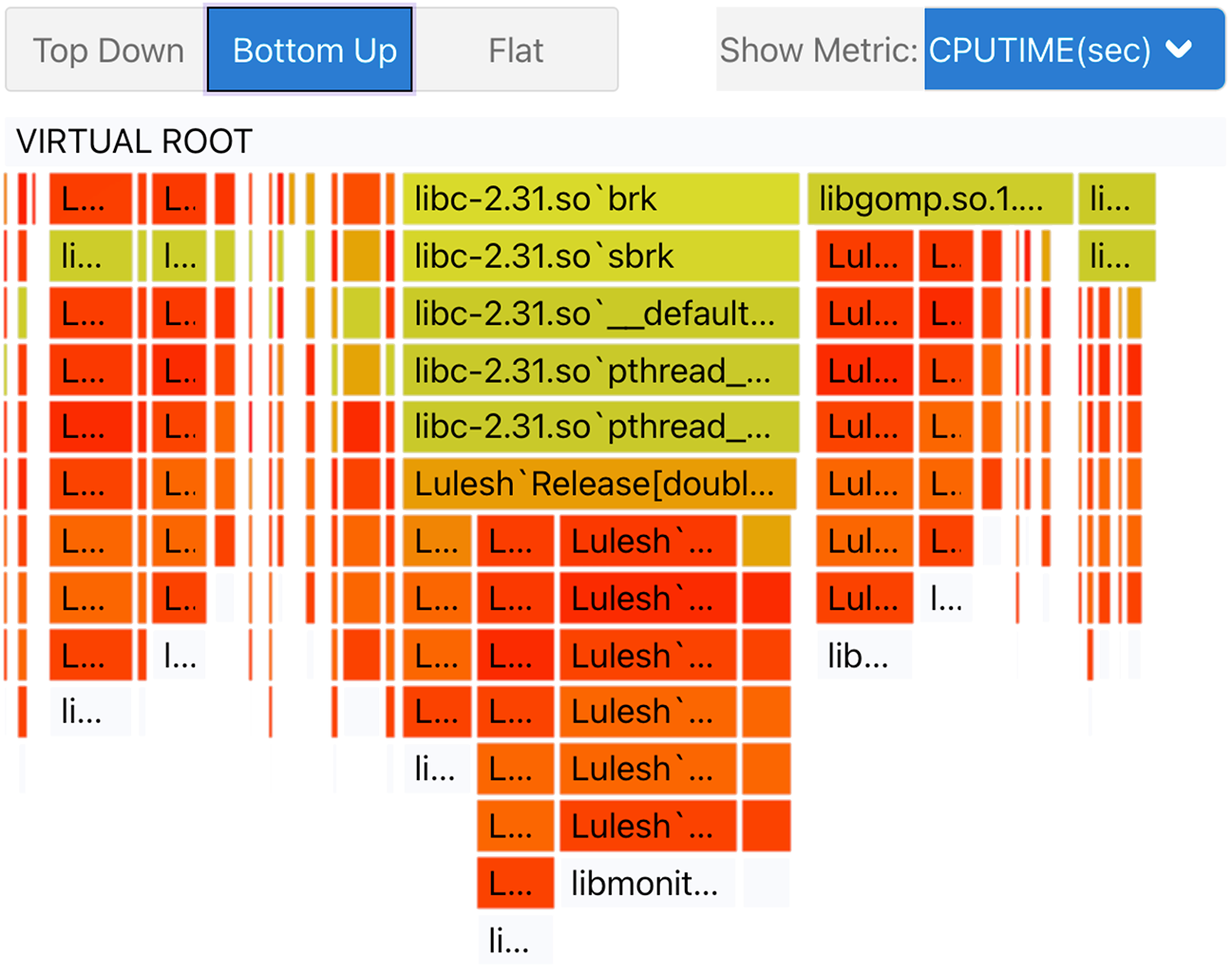}
  \end{center}
  \caption{The bottom-up flame graph presented by \tool{} for LULESH. The profile, collected by HPCToolkit, shows CPU time as the metric. A frame is denoted as the library name + function name. A large frame means a hotspot in the code. The depth is the reversed call path (top is the leaf function and bottom is the main function). The function {\tt brk} in {\tt libc-2.31.so} is a clear hotspot in the execution.} 
  \label{fig:lulesh-h-bu}
\end{figure}

\begin{figure*}[h]
\centering
\includegraphics[width=.88\textwidth]{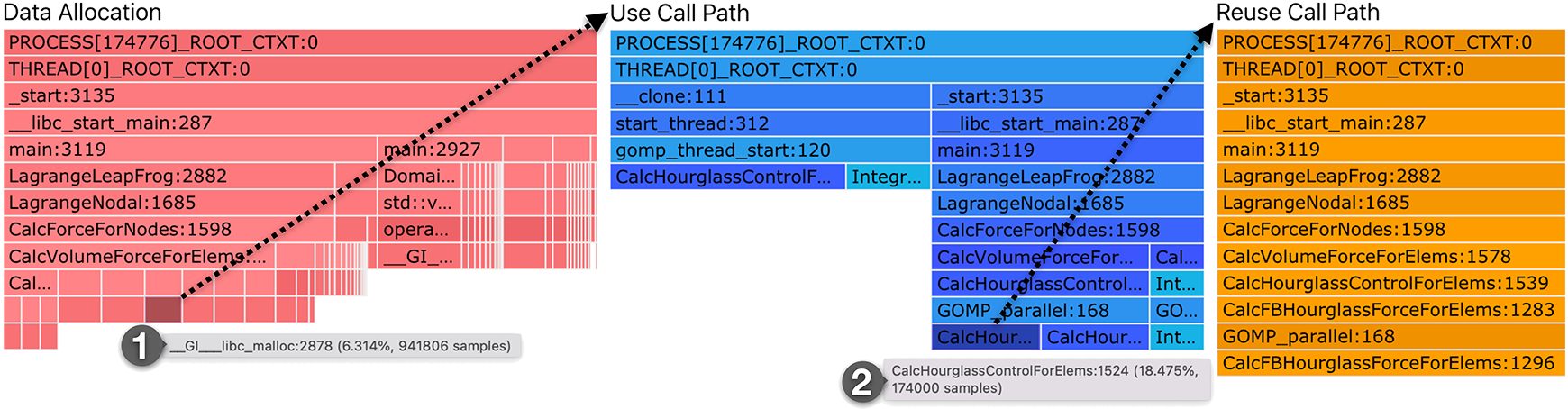}
\centering
\caption{\tool{} shows the LULESH profile collected by DrCCTProf. There are multiple flame graphs correlated with each other to show the data allocations (\textcircled{1}) and their use-reuse pairs in their call paths (\textcircled{2}). This snapshot highlights the reuses that are optimizable in the hotspot function shown in the top-down view (not shown in the paper). }
\vspace{-0.1in}
\label{fig:lulesh-d-ru}
\end{figure*}

In the HPC Domain, we show how \tool{} combines the outputs of  two mainstream profilers to analyze LULESH~\cite{LULESH}, a proxy application written in C++ developed by Lawrence Livermore National Laboratory to solve the Sedov blast wave problem for one material in three dimensions. The two profilers are HPCToolkit~\cite{adhianto2010hpctoolkit} and DrCCTProf~\cite{drcctprof}, which are state-of-the-art profilers for HPC applications.
We use them for complementary insights: HPCToolkit pinpoints hotspots that are worthy of further investigation;, while DrCCTProf identifies some root causes of inefficiencies. We show that \tool{} is able to combine the two independent profiles in a unified view for more intuitive analyses.

\tool{} first shows the flame graphs for hotspot analysis measured by HPCToolkit. 
Figure~\ref{fig:lulesh-h-bu} is a bottom-up flame graph, which shows the hot leaf function in the reversed call paths. It is straightforward to see that function {\tt brk} from the {\tt libc.so} library is the hotspot, which is called in multiple call paths. With further investigation of these reversed call paths and their attribution to the source code, we can see that the hotspot is rooted in the memory management (i.e., memory allocation and release). We replace the default memory management in {\tt libc} with {\tt TCMalloc}~\cite{tcmalloc}, a more efficient solution, which yields a 30\% speedup to the entire program.

\sloppy
The top-down flame graph (not shown in the paper) highlights the function {\tt CalcVolumeForceForElems} and its callee {\tt  CalcHourglassForceForElems} as hotspots. To understand the root causes, we investigate the profile produced by DrCCTProf. DrCCTProf measures the memory reuses~\cite{drcctprof} to quantify the data locality in LULESH. As shown in Figure~\ref{fig:lulesh-d-ru}, three flame graphs are correlated in \tool{}: the left one shows all the array allocations; when selecting one array allocation (\textcircled{1}), the middle one appears to show all the uses to this array. When selecting a use (\textcircled{2}), the right one appears to show all the reuses following this use. The metric is the memory bytes allocated  for the left flame graph and occurrences of memory accesses for the other two flame graphs. Figure~\ref{fig:lulesh-d-ru} shows a reuse tuple related to the hot function {\tt  CalcHourglassForceForElems}. \tool{} correlates the three flame graphs and clearly shows the call paths, which can easily guide locality optimization: hoisting the use and reuse code to the least common ancestor of the call paths and performing the loop fusion. The optimization yields an additional 28\% speedup to the entire LULESH program. 

%

\textit{Comparing to off-the-shelf HPCToolkit and DrCCTProf:} We find that their original GUIs are not as straightforward as \tool{} and  require substantial efforts to learn and use to obtain the same insights compared to \tool{}. Moreover, these two tools have their own GUIs that are separated from the IDEs, which cannot easily combine their profiles in a unified view for easy analysis. 

\subsection{User Studies}
\label{user}

To give a fair evaluation of \tool{} on helping software engineers analyze their software, we perform a user study in a broader group beyond the graduate students. 
We released \tool{} in VSCode market and recorded a few short videos for the tutorial to help users install and use \tool{}. We sent emails to the mailing lists and posted them on the technique forums to recruit potential users. We invite programmers with at least 1-year programming experience to use \tool{}, regardless of their experiences in code profiling and optimization. After trying out \tool{}, users are invited to fill out an anonymous survey form. By the paper submission, we found 216 installations of \tool{} recorded in VSCode marketplace and received 26 completed forms. We show the statistics of the survey as follows.

We first let the survey participants identify their own expertise in code profiling and optimization to ensure this survey covers programmers of different levels. Among them, 53.8\% actively tune the code for high performance. There are 34.6\%, 19.2\%, and 46.2\% of participants using profilers at the frequency of weekly, monthly, and very rarely, respectively. 
Among all the participants, 92.3\% agree that \tool{}, integrating the profile analysis and visualization within VSCode,  is effective in facilitating performance analysis. It is worth noting that even the participants with limited knowledge in performance analysis agree \tool{} is useful, which flattens their learning curves of performance analysis. 7.7\% of participants question the effectiveness of \tool{}; they are not VSCode users as they suggest \tool{} support other IDEs besides VSCode.
We further ask the participants to quantify the performance improvement of their code guided by \tool{}. 
80.8\% of participants improve their code performance with \tool{}'s help. The speedups range from 1.03$\times$ to 3$\times$ from optimizing hotspot, memory, and mutex. For the participants who do not obtain performance improvement, they also admit that \tool{} helps pinpoint performance inefficiencies worth of further investigation.

\begin{figure}[t]
  \begin{center}
  \includegraphics[width=0.45\textwidth]{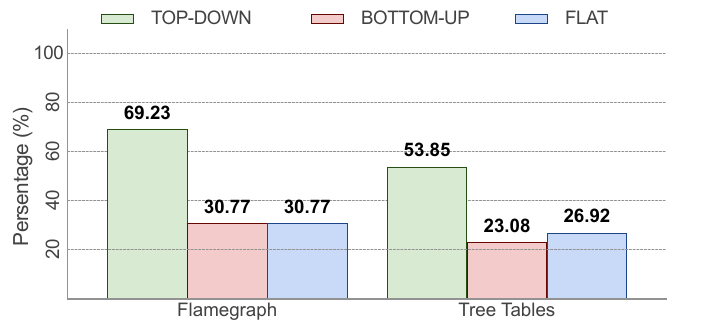}
  \end{center}
  \caption{Comparing the effectiveness of different views. The y-axis quantifies the percentage of all participants that believe the views are effective. Each participant can specify zero or multiple choices.} 
  \vspace{-0.1in}
  \label{fig:userstudy-c}
\end{figure}

Our user studies also show the comparison among different views in Figure~\ref{fig:userstudy-c}. Overall, flame graphs are more effective than tree tables (92.3\% vs. 84.6\%). Among all of the views, the top-down views are the most helpful, compared to bottom-up and flat views in both flame graphs and tree tables. These insights guide the evolution of \tool{} in two directions: (1) deriving various analyses based on the top-down view and (2) demonstrating more use cases for bottom-up and flat views to enable their wide adoption.

\textbf{Control group evaluation.} To further compare \tool{} with existing approaches, we create experimental and control groups to analyze the performance data collected by PProf. The experimental group (7 people) use \tool{}, while two control groups: 7 people use the default PProf visualizer and 7 people use JetBrain's GoLand PProf plugin. People in different groups are mixed with newbies and experienced engineers in performance engineering, all of whom are trained with the basic knowledge about flame graphs~\cite{flamegraph}. All the groups are given the same set of profiles and are asked to perform three tasks. \emph{Task I:} Pinpointing hotspot functions in their calling contexts for high CPU and memory consumption. This is a  use case for top-down flame graphs. \emph{Task II:} Identifying hot memory allocation, garbage collection invocation, and lock wait, and figuring out where they are called. This is a  use case for bottom-up flame graphs. \emph{Task III:} Identifying memory leak as described in Section~\ref{cloud}. This is to evaluate handling multiple profiles. The observation is as follows.

\begin{itemize}[leftmargin=*]
\item For Task I, the experimental group uses $\sim$10 min on average to investigate all the profiles. The control group using GoLand spends longer time ($\sim$15 min). Although the top-down flame graph produced by \tool{} and GoLand are similar, GoLand requires much more time to open and navigate large profiles. The control group using default PProf requires most time ($sim$30min) to complete the task. It is mainly because unlike \tool{} and GoLand, PProf requires manual correlate profiles with source code. 

\item For Task II, the experimental group uses $\sim$10 min on average to complete. The control group using GoLand needs $\sim$1 hour as GoLand does not provide bottom-up flame graphs for analysis. Instead, it provides a bottom-up tree table, which requires more learning time. The control group using PProf needs more than 3 hours to complete the task because PProf does not provide any bottom-up view but requires tedious manual analysis. 

\item For Task III, the experimental group uses $\sim$10 min to complete. Both control groups cannot complete the task in 3 hours. It is because PProf and GoLand do not provide an easy way to analyze multiple profiles. One needs to either analyze multiple profiles manually or devise a script for automatic analysis, both of which add significant burdens to users.

\end{itemize}

In summary, \tool{} significantly flattens the learning curves for various performance analyses. Because \tool{} integrate these analyses in a consistent flame graph view, one needs to spend minimum efforts to master it.



\section{Threats to validity}
\label{subsec:threat}

Like all existing approaches based on empirical and user studies, our research suffers from internal and external threats.

\paragraph{Internal threats}

We suffer from the internal threats in picking up the representative survey participants, which may result in biased results. 
The participants may have different levels of knowledge in performance tuning with profilers, different levels of expertise in the use of Microsoft VSCode, different levels of familiarity in flame graphs or tree tables. 
To obtain representative results, we use three manners to minimize the internal threats. First, we disseminate our surveys via multiple channels (e.g., mailing list, tech forums, and community slack/discord) to invite a relatively large number of participants. Second, we pre-record several short videos to demonstrate the usage of \tool{}. Users can use these videos as the tutorials to gain the basic knowledge of \tool{}. Third, we actively answer questions raised by the users to ensure they can use \tool{} correctly and efficiently. In our survey, we explicitly ask the users' programming knowledge and find that our survey is representative, covering programmers of different levels. 

Another internal threat is related to the claim that \tool{} provides general support for profilers. We are not able to study all profilers in practice, but we guarantee our claim applies to the popular and mainstream profilers as described in Section~\ref{representation}.

\paragraph{External threats}
We identify one potential external threat that we use a Google form to maintain all the survey questions. Users of \tool{} may not actively fill the form, as participating the survey can distract them from their daily work. To increase the response ratio, we highlight the survey link in \tool{} webpage in the VSCode marketplace.
Moreover, \tool{} is currently released as an extension of VSCode only. It excludes some users that actively use other IDEs, such as IntelliJ or Eclipse. Supporting other popular IDEs is under development.

\section{Conclusions}
\label{sec:conclusion}

In this paper, we describe \tool{}, which brings performance profiles into IDEs to facilitate performance analysis for both deeper insights and better user experiences. \tool{} employs a generic profiling data representation that covers all basic features and many advanced features of mainstream profilers. Moreover, \tool{} unifies the data analysis with predefined schemes as well as customized schemes for one or multiple profiles. \tool{} visualizes the performance data in variants of flame graphs and tree tables, which are tightly integrated into IDEs with low response time. With our empirical studies, we show \tool{} provides simple APIs to support various tools and deep insights to analyze workloads in different domains. We further leverage survey forms for user studies. A majority of the participants agree that \tool{} can significantly help their efforts in understanding program performance.  Many users have obtained nontrivial speedups with the optimization guided by \tool{}.  Our controlled experiments show that \tool{} significantly outperforms state of the arts in interpreting profiling data.


\section*{Acknowledgements}
We thank anonymous reviewers for their valuable comments. This research is partially supported by NSF CNS 2050007 and a Google gift.

\appendix
\section{Artifact Appendix}

\subsection{Abstract}

The provided artifact encompasses an extension of the Visual Studio Code, specifically engineered for the purpose of visualization evaluation. Additionally, it includes a binary executable file used to evaluate response time and a Python script for Profile converter. The artifact is published in Zenodo~\cite{easyview-ae}.

\subsection{Artifact check-list (meta-information)}

{\small
\begin{itemize}
  \item {\bf Program: }
  The artifact includes a profile format converter, which has been implemented using the Python programming language.
  \item {\bf Binary: }
  The artifact comprises a Visual Studio Code Extension and an executed binary file.
  \item {\bf Hardware: }
  The physical machine must have at least 8GB of memory, and the CPU architecture should be either Intel x86-64 or ARM64.
  \item {\bf Data set: }
  The artifact includes three overhead evaluation profiles.
  \item {\bf How much disk space required (approximately)?: }
   10 GB
  \item {\bf How much time is needed to prepare workflow (approximately)?: }
  10 minutes
  \item {\bf How much time is needed to complete experiments (approximately)?: }
  1 hour
  \item {\bf Publicly available?: }
  The profile format converter in the artifact is open-source.
\end{itemize}

\subsection{Installation}
The installation involves three key steps: firstly, installing the background Docker container; secondly, connecting Visual Studio Code to the Docker container; and thirdly, installing the Visual Studio Code extension.

\subsubsection{\bf Docker Container Installation}

\begin{itemize}
\item {To install the Docker container, open a terminal and enter the command below. }
\begin{lstlisting}[language=bash]
# On Intel machine
$docker pull qzhao24/easyview-cgo23
$docker run --hostname=easyview-cgo23 -d -p 2222:22 qzhao24/easyview-cgo23
# On Arm machine
$docker pull qzhao24/easyview-cgo23-arm64
$docker run --hostname=easyview-cgo23 -d -p 2222:22 qzhao24/easyview-cgo23-arm64
\end{lstlisting}
\vspace{-\baselineskip}
\end{itemize}

\subsubsection{\bf Connecting to Docker Container in Visual Studio Code}

\begin{itemize}
\item {Install Remote Development Extension~\cite{vscode-remote-ex}. }
\item {Use the extension to connect the docker container. The SSH command is as follows}
\begin{lstlisting}[language=bash]
# The password is '1234'
$ssh -p 2222 qzhao24@localhost
\end{lstlisting}
\vspace{-\baselineskip}
\end{itemize}

\subsubsection{\bf Visual Studio Code Extension Installation}
\begin{itemize}
\item {After establishing a connection to the Docker container, proceed to install the Visual Studio Code extension using a .vsix file located within the Docker container:}
\begin{lstlisting}[language=bash]
/home/qzhao24/easyview-artifact/Extension/easyviewcgo23-0.2.0.vsix
\end{lstlisting}
\vspace{-\baselineskip}
\item {Access the extension window, locate the item "EasyViewCGO23," and click trust it on its detail window.
}
\end{itemize}

\subsection{Experiment workflow}

This experiment is structured into three parts:
\subsubsection{\bf Profile Format Converter Evaluation}

\begin{itemize}
\item {To conduct the experiment provided by this artifact, follow these steps:
First, open a terminal within Visual Studio Code, ensuring it is connected to the Docker container. Then, execute the command provided below in the terminal.}

\begin{lstlisting}[language=bash]
$cd /home/qzhao24/easyview-artifact/Source/drcctprof-databuilder
$./hpctoolkit-converter.py /home/qzhao24/easyview-artifact/Source/hpctoolkit-lulesh-par-original-database lulesh.hpctoolkit.drcctprof
\end{lstlisting}
\vspace{-\baselineskip}
\end{itemize}

\subsubsection{\bf Response Time Assessment}

\begin{itemize}
\item {To conduct the experiment provided by this artifact, follow these steps:
First, open a terminal within Visual Studio Code, ensuring it is connected to the Docker container. Then, execute the command provided below in the terminal.}
\begin{lstlisting}[language=bash]
$cd /home/qzhao24/easyview-artifact/Overhead
# On Intel machine
$./easyview_overhead_test 1M.profile
$./easyview_overhead_test 100M.profile
$./easyview_overhead_test 1GB.profile
# On Arm machine
$./easyview_overhead_test_arm64 1M.profile
$./easyview_overhead_test_arm64 100M.profile
$./easyview_overhead_test_arm64 1GB.profile
\end{lstlisting}
\vspace{-\baselineskip}

\end{itemize}

\subsubsection{\bf Virtualization Evaluation}

\begin{itemize}
\item {To conduct the experiment associated with this artifact, employ Visual Studio Code to access the specified files:}

\begin{lstlisting}[language=bash]
$/home/qzhao24/easyview-artifact/Profiles/fig4_memory_flow.ezview
$/home/qzhao24/easyview-artifact/Profiles/fig6_lulesh_bottomup.ezview
$/home/qzhao24/easyview-artifact/Profiles/fig7_lulesh_reuse_distance.ezview
\end{lstlisting}
\vspace{-\baselineskip}
\end{itemize}

\subsection{Evaluation and expected result}

Regarding the three experiments described:

\subsubsection{\bf Profile Format Converter Evaluation}

\begin{itemize}
\item {The anticipated outcome of this evaluation is the successful execution of the command, resulting in the generation of a .drcctprof file. Subsequently, this profile should be accessible via Visual Studio Code, allowing for the display of a Flamegraph.}
\end{itemize}

\subsubsection{\bf Response time assessment}

\begin{itemize}
\item {The expected outcomes involve the successful execution of the last three commands, resulting in the generation of execution time outputs.}
\end{itemize}

\subsubsection{\bf Virtualization evaluation}

\begin{itemize}
\item {The expected results for this evaluation involve the three profiles showing a figure that closely resembles the one presented in the paper.}
\end{itemize}

\balance{}
\bibliography{IEEEabrv,references}

\begin{thebibliography}{10}
\providecommand{\url}[1]{#1}
\csname url@samestyle\endcsname
\providecommand{\newblock}{\relax}
\providecommand{\bibinfo}[2]{#2}
\providecommand{\BIBentrySTDinterwordspacing}{\spaceskip=0pt\relax}
\providecommand{\BIBentryALTinterwordstretchfactor}{4}
\providecommand{\BIBentryALTinterwordspacing}{\spaceskip=\fontdimen2\font plus
\BIBentryALTinterwordstretchfactor\fontdimen3\font minus
  \fontdimen4\font\relax}
\providecommand{\BIBforeignlanguage}[2]{{%
\expandafter\ifx\csname l@#1\endcsname\relax
\typeout{** WARNING: IEEEtran.bst: No hyphenation pattern has been}%
\typeout{** loaded for the language `#1'. Using the pattern for}%
\typeout{** the default language instead.}%
\else
\language=\csname l@#1\endcsname
\fi
#2}}
\providecommand{\BIBdecl}{\relax}
\BIBdecl

\bibitem{Molyneaux:2009:AAP:1550832}
I.~Molyneaux, \emph{The Art of Application Performance Testing: Help for
  Programmers and Quality Assurance}, 1st~ed.\hskip 1em plus 0.5em minus
  0.4em\relax O'Reilly Media, Inc., 2009.

\bibitem{Bryant:2010:CSP:1841497}
R.~E. Bryant and D.~R. O'Hallaron, \emph{Computer Systems: A Programmer's
  Perspective}, 2nd~ed.\hskip 1em plus 0.5em minus 0.4em\relax USA:
  Addison-Wesley Publishing Company, 2010.

\bibitem{deitz2001eliminating}
S.~J. Deitz, B.~L. Chamberlain, and L.~Snyder, ``{Eliminating Redundancies in
  Sum-of-product Array Computations},'' in \emph{Proceedings of the 15th
  International Conference on Supercomputing}, ser. ICS '01.\hskip 1em plus
  0.5em minus 0.4em\relax New York, NY, USA: ACM, 2001, pp. 65--77.

\bibitem{gvn}
B.~K. Rosen, M.~N. Wegman, and F.~K. Zadeck, ``{Global Value Numbers and
  Redundant Computations},'' in \emph{Proceedings of the 15th ACM
  SIGPLAN-SIGACT Symposium on Principles of Programming Languages}, 1988, pp.
  12--27.

\bibitem{Wegman:1991:CPC:103135.103136}
M.~N. Wegman and F.~K. Zadeck, ``{Constant Propagation with Conditional
  Branches},'' \emph{ACM Trans. Program. Lang. Syst.}, vol.~13, no.~2, pp.
  181--210, Apr 1991.

\bibitem{adhianto2010hpctoolkit}
L.~Adhianto, S.~Banerjee, M.~Fagan, M.~Krentel, G.~Marin, J.~Mellor-Crummey,
  and N.~R. Tallent, ``{HPCToolkit: Tools for Performance Analysis of Optimized
  Parallel Programs},'' \emph{Concurrency Computation : Practice Expererience},
  vol.~22, no.~6, pp. 685--701, Apr 2010.

\bibitem{vtune}
``{Intel VTune},''
  \url{https://software.intel.com/en-us/intel-vtune-amplifier-xe}, 2018.

\bibitem{Graham-etal:1982:PLDI-gprof}
S.~L. Graham, P.~B. Kessler, and M.~K. Mckusick, ``{Gprof: A Call Graph
  Execution Profiler},'' in \emph{Proceedings of the 1982 SIGPLAN Symposium on
  Compiler Construction}, ser. SIGPLAN '82.\hskip 1em plus 0.5em minus
  0.4em\relax New York, NY, USA: ACM, 1982, pp. 120--126.

\bibitem{pprof}
G.~Inc., ``{PProf},'' \url{https://github.com/google/pprof}.

\bibitem{Levon:OProfile}
J.~{Levon \textit{et al.}}, ``{OProfile},''
  \url{http://oprofile.sourceforge.net}, 2017.

\bibitem{perf}
Linux, ``Linux perf tool,''
  \url{https://perf.wiki.kernel.org/index.php/Main_Page}, 2015.

\bibitem{10.1109/TSE.2006.116}
\BIBentryALTinterwordspacing
A.~J. Ko, B.~A. Myers, M.~J. Coblenz, and H.~H. Aung, ``An exploratory study of
  how developers seek, relate, and collect relevant information during software
  maintenance tasks,'' \emph{IEEE Trans. Softw. Eng.}, vol.~32, no.~12, p.
  971–987, Dec. 2006. [Online]. Available:
  \url{https://doi.org/10.1109/TSE.2006.116}
\BIBentrySTDinterwordspacing

\bibitem{interrupt}
C.~Parnin and S.~Rugaber, ``Resumption strategies for interrupted programming
  tasks,'' in \emph{Software Quality Journal}, 2011.

\bibitem{csikszentmihalyi2014flow}
M.~Csikszentmihalyi and R.~Larson, \emph{Flow and the foundations of positive
  psychology:The Collected Works of Mihaly Csikszentmihalyi}.\hskip 1em plus
  0.5em minus 0.4em\relax Springer, 2014, vol.~10.

\bibitem{10.5555/951952.952340}
D.~Saff and M.~D. Ernst, ``Reducing wasted development time via continuous
  testing,'' in \emph{Proceedings of the 14th International Symposium on
  Software Reliability Engineering}, ser. ISSRE '03.\hskip 1em plus 0.5em minus
  0.4em\relax USA: IEEE Computer Society, 2003, p. 281.

\bibitem{magpiebridge}
\BIBentryALTinterwordspacing
L.~Luo, J.~Dolby, and E.~Bodden, ``{MagpieBridge: A General Approach to
  Integrating Static Analyses into IDEs and Editors (Tool Insights Paper)},''
  in \emph{33rd European Conference on Object-Oriented Programming (ECOOP
  2019)}, ser. Leibniz International Proceedings in Informatics (LIPIcs), A.~F.
  Donaldson, Ed., vol. 134.\hskip 1em plus 0.5em minus 0.4em\relax Dagstuhl,
  Germany: Schloss Dagstuhl--Leibniz-Zentrum fuer Informatik, 2019, pp.
  21:1--21:25. [Online]. Available:
  \url{http://drops.dagstuhl.de/opus/volltexte/2019/10813}
\BIBentrySTDinterwordspacing

\bibitem{xforce}
I.~Inc., ``{IBM AppScan},'' \url{https://www.ibm.com/security}.

\bibitem{whitesource}
``{The Python Profilers},''
  \url{https://docs.python.org/3/library/profile.html}.

\bibitem{lsp}
M.~Inc., ``{Language Server Protocol},''
  \url{https://microsoft.github.io/language-server-protocol/}.

\bibitem{async-profiler}
``{async-profiler},''
  \url{https://github.com/jvm-profiling-tools/async-profiler}.

\bibitem{tau}
U.~of~Oregon, ``{TAU: Tuning and Analysis Utilities},''
  \url{https://www.cs.uoregon.edu/research/tau/home.php}.

\bibitem{dynatrace}
``{Dynatrace},'' \url{https://www.dynatrace.com}.

\bibitem{vs}
M.~Inc., ``{Visual Studio},'' \url{https://visualstudio.microsoft.com}.

\bibitem{jetbrain}
J.~s.r.o., ``{JetBrains},'' \url{https://www.jetbrains.com}.

\bibitem{eclipse}
E.~Foundation, ``{Eclipse},'' \url{https://www.eclipse.org}.

\bibitem{ICPC-2013-BeckMDR}
F.~Beck, O.~Moseler, S.~Diehl, and G.~D. Rey, ``{In situ understanding of
  performance bottlenecks through visually augmented code},'' in
  \emph{{Proceedings of the 21st International Conference on Program
  Comprehension}}, 2013, pp. 63--72.

\bibitem{vscode}
M.~Inc., ``{Visual Studio Code},'' \url{https://code.visualstudio.com}.

\bibitem{loadspy}
P.~Su, S.~Wen, H.~Yang, M.~Chabbi, and X.~Liu, ``Redundant loads: A software
  inefficiency indicator,'' in \emph{2019 IEEE/ACM 41st International
  Conference on Software Engineering (ICSE)}, May 2019.

\bibitem{jxperf}
\BIBentryALTinterwordspacing
P.~Su, Q.~Wang, M.~Chabbi, and X.~Liu, ``Pinpointing performance inefficiencies
  in java,'' in \emph{Proceedings of the 2019 27th ACM Joint Meeting on
  European Software Engineering Conference and Symposium on the Foundations of
  Software Engineering}, ser. ESEC/FSE 2019.\hskip 1em plus 0.5em minus
  0.4em\relax New York, NY, USA: Association for Computing Machinery, 2019, p.
  818–829. [Online]. Available: \url{https://doi.org/10.1145/3338906.3338923}
\BIBentrySTDinterwordspacing

\bibitem{valgrind}
\BIBentryALTinterwordspacing
N.~Nethercote and J.~Seward, ``Valgrind: A framework for heavyweight dynamic
  binary instrumentation,'' in \emph{Proceedings of the 28th ACM SIGPLAN
  Conference on Programming Language Design and Implementation}, ser. PLDI
  '07.\hskip 1em plus 0.5em minus 0.4em\relax New York, NY, USA: Association
  for Computing Machinery, 2007, p. 89–100. [Online]. Available:
  \url{https://doi.org/10.1145/1250734.1250746}
\BIBentrySTDinterwordspacing

\bibitem{berger2020scalene}
E.~D. Berger, ``Scalene: Scripting-language aware profiling for python,''
  \emph{arXiv preprint arXiv:2006.03879}, 2020.

\bibitem{markdown}
``{Basic Syntax of Markdown Languages},''
  \url{https://www.markdownguide.org/basic-syntax/}.

\bibitem{hotspot}
``{Hotspot - the Linux perf GUI for performance analysis},''
  \url{https://github.com/KDAB/hotspot}, 2022.

\bibitem{googleprofiler}
``{Google Cloud Profiler},'' \url{https://cloud.google.com/profiler}.

\bibitem{perffeto}
``{Perffeto: System profiling, app tracing and trace analysis},''
  \url{https://perfetto.dev}.

\bibitem{speedscope}
``{SpeedScope},'' \url{https://www.speedscope.app}.

\bibitem{Pyroscope}
``{Pyroscope: Open Source Continuous Profiling Platform},''
  \url{https://pyroscope.io}.

\bibitem{gProfiler}
``{gProfiler},'' \url{https://gprofiler.io/}.

\bibitem{Flamescope}
``{Flamescope},'' \url{https://github.com/Netflix/flamescope}.

\bibitem{intellij}
``{Profiler in IntelliJ},''
  \url{https://www.jetbrains.com/help/idea/cpu-profiler.html}.

\bibitem{clion}
``{Profiler in CLion},''
  \url{https://www.jetbrains.com/help/clion/cpu-profiler.html}.

\bibitem{goland}
``{Profiler in Goland},''
  \url{https://www.jetbrains.com/help/go/profiling-tests-and-benchmarks.html}.

\bibitem{vtune-vscode}
``{Using Intel VTune Profiler Server with Visual Studio Code and Intel DevCloud
  for oneAPI},''
  \url{https://www.intel.com/content/www/us/en/develop/documentation/vtune-cookbook/top/configuration-recipes/using-vtune-server-with-vs-code-intel-devcloud.html}.

\bibitem{pprof-vscode}
\url{https://marketplace.visualstudio.com/items?itemName=MaxMedia.go-prof}.

\bibitem{austin-vscode}
``{Austin VS Code Extension},''
  \url{https://github.com/p403n1x87/austin-vscode}.

\bibitem{performancehat}
\BIBentryALTinterwordspacing
J.~Cito, P.~Leitner, C.~Bosshard, M.~Knecht, G.~Mazlami, and H.~C. Gall,
  ``Performancehat: Augmenting source code with runtime performance traces in
  the ide,'' in \emph{Proceedings of the 40th International Conference on
  Software Engineering: Companion Proceeedings}, ser. ICSE '18.\hskip 1em plus
  0.5em minus 0.4em\relax New York, NY, USA: Association for Computing
  Machinery, 2018, p. 41–44. [Online]. Available:
  \url{https://doi.org/10.1145/3183440.3183481}
\BIBentrySTDinterwordspacing

\bibitem{10.1145/3173574.3174106}
\BIBentryALTinterwordspacing
J.~Hoffswell, A.~Satyanarayan, and J.~Heer, \emph{Augmenting Code with In Situ
  Visualizations to Aid Program Understanding}.\hskip 1em plus 0.5em minus
  0.4em\relax New York, NY, USA: Association for Computing Machinery, 2018, p.
  1–12. [Online]. Available: \url{https://doi.org/10.1145/3173574.3174106}
\BIBentrySTDinterwordspacing

\bibitem{10.1145/2556288.2557409}
\BIBentryALTinterwordspacing
T.~Lieber, J.~R. Brandt, and R.~C. Miller, ``Addressing misconceptions about
  code with always-on programming visualizations,'' in \emph{Proceedings of the
  SIGCHI Conference on Human Factors in Computing Systems}, ser. CHI '14.\hskip
  1em plus 0.5em minus 0.4em\relax New York, NY, USA: Association for Computing
  Machinery, 2014, p. 2481–2490. [Online]. Available:
  \url{https://doi.org/10.1145/2556288.2557409}
\BIBentrySTDinterwordspacing

\bibitem{scorep}
A.~Kn{\"u}pfer, C.~R{\"o}ssel, D.~a. Mey, S.~Biersdorff, K.~Diethelm,
  D.~Eschweiler, M.~Geimer, M.~Gerndt, D.~Lorenz, A.~Malony, W.~E. Nagel,
  Y.~Oleynik, P.~Philippen, P.~Saviankou, D.~Schmidl, S.~Shende,
  R.~Tsch{\"u}ter, M.~Wagner, B.~Wesarg, and F.~Wolf, ``Score-p: A joint
  performance measurement run-time infrastructure for periscope,scalasca, tau,
  and vampir,'' in \emph{Tools for High Performance Computing 2011}, H.~Brunst,
  M.~S. M{\"u}ller, W.~E. Nagel, and M.~M. Resch, Eds.\hskip 1em plus 0.5em
  minus 0.4em\relax Berlin, Heidelberg: Springer Berlin Heidelberg, 2012, pp.
  79--91.

\bibitem{caliper}
D.~Boehme, T.~Gamblin, D.~Beckingsale, P.-T. Bremer, A.~Gimenez, M.~LeGendre,
  O.~Pearce, and M.~Schulz, ``Caliper: Performance introspection for hpc
  software stacks,'' in \emph{SC '16: Proceedings of the International
  Conference for High Performance Computing, Networking, Storage and Analysis},
  2016, pp. 550--560.

\bibitem{armmap}
L.~Limited, ``{Linaro MAP},'' \url{https://www.linaroforge.com/linaroMap}.

\bibitem{uprof}
A.~Inc., ``{AMD uProf},'' \url{https://developer.amd.com/amd-uprof}.

\bibitem{nsight}
N.~Inc., ``{NVIDIA Nsight Compute},''
  \url{https://developer.nvidia.com/nsight-compute}.

\bibitem{cctlib}
\BIBentryALTinterwordspacing
M.~Chabbi, X.~Liu, and J.~Mellor-Crummey, ``Call paths for pin tools,'' in
  \emph{Proceedings of Annual IEEE/ACM International Symposium on Code
  Generation and Optimization}, ser. CGO '14.\hskip 1em plus 0.5em minus
  0.4em\relax New York, NY, USA: Association for Computing Machinery, 2014, p.
  76–86. [Online]. Available: \url{https://doi.org/10.1145/2581122.2544164}
\BIBentrySTDinterwordspacing

\bibitem{drcctprof}
Q.~Zhao, X.~Liu, and M.~Chabbi, ``Drcctprof: A fine-grained call path profiler
  for arm-based clusters,'' in \emph{Proceedings of the International
  Conference for High Performance Computing, Networking, Storage and Analysis},
  ser. SC '20.\hskip 1em plus 0.5em minus 0.4em\relax IEEE Press, 2020.

\bibitem{Luk:2005:PBC:1065010.1065034}
C.-K. Luk, R.~Cohn, R.~Muth, H.~Patil, A.~Klauser, G.~Lowney, S.~Wallace, V.~J.
  Reddi, and K.~Hazelwood, ``Pin: Building customized program analysis tools
  with dynamic instrumentation,'' in \emph{Proceedings of the 2005 ACM SIGPLAN
  Conference on Programming Language Design and Implementation}, ser. PLDI
  '05.\hskip 1em plus 0.5em minus 0.4em\relax New York, NY, USA: ACM, 2005, pp.
  190--200.

\bibitem{nvbit}
\BIBentryALTinterwordspacing
O.~Villa, M.~Stephenson, D.~Nellans, and S.~W. Keckler, ``Nvbit: A dynamic
  binary instrumentation framework for nvidia gpus,'' in \emph{Proceedings of
  the 52nd Annual IEEE/ACM International Symposium on Microarchitecture}, ser.
  MICRO '52.\hskip 1em plus 0.5em minus 0.4em\relax New York, NY, USA:
  Association for Computing Machinery, 2019, p. 372–383. [Online]. Available:
  \url{https://doi.org/10.1145/3352460.3358307}
\BIBentrySTDinterwordspacing

\bibitem{proto}
G.~Inc., ``{Protocol Buffers},''
  \url{https://developers.google.com/protocol-buffers}.

\bibitem{Liu:2015:STI:2807591.2807648}
\BIBentryALTinterwordspacing
X.~Liu and B.~Wu, ``Scaanalyzer: A tool to identify memory scalability
  bottlenecks in parallel programs,'' in \emph{Proceedings of the International
  Conference for High Performance Computing, Networking, Storage and Analysis},
  ser. SC '15.\hskip 1em plus 0.5em minus 0.4em\relax New York, NY, USA: ACM,
  2015, pp. 47:1--47:12. [Online]. Available:
  \url{http://doi.acm.org/10.1145/2807591.2807648}
\BIBentrySTDinterwordspacing

\bibitem{Liu:2016:CDF:2854038.2854039}
\BIBentryALTinterwordspacing
T.~Liu and X.~Liu, ``Cheetah: Detecting false sharing efficiently and
  effectively,'' in \emph{Proceedings of the 2016 International Symposium on
  Code Generation and Optimization}, ser. CGO '16.\hskip 1em plus 0.5em minus
  0.4em\relax New York, NY, USA: ACM, 2016, pp. 1--11. [Online]. Available:
  \url{http://doi.acm.org/10.1145/2854038.2854039}
\BIBentrySTDinterwordspacing

\bibitem{memprof}
R.~Lachaize, B.~Lepers, and V.~Qu{\'e}ma, ``{MemProf}: A memory profiler for
  {NUMA} multicore systems,'' in \emph{USENIX ATC}, 2012.

\bibitem{Marin-sweep3d}
G.~Marin and J.~Mellor-Crummey, ``{Pinpointing and Exploiting Opportunities for
  Enhancing Data Reuse},'' in \emph{IEEE Intl. Symposium on Performance
  Analysis of Systems and Software}, ser. ISPASS '08.\hskip 1em plus 0.5em
  minus 0.4em\relax Washington, DC, USA: IEEE Computer Society, 2008, pp.
  115--126.

\bibitem{redspy}
S.~Wen, M.~Chabbi, and X.~Liu, ``Redspy: Exploring value locality in
  software,'' in \emph{Proceedings of the Twenty-Second International
  Conference on Architectural Support for Programming Languages and Operating
  Systems}, ser. ASPLOS '17.\hskip 1em plus 0.5em minus 0.4em\relax New York,
  NY, USA: ACM, 2017, pp. 47--61.

\bibitem{witch}
S.~Wen, X.~Liu, J.~Byrne, and M.~Chabbi, ``Watching for software inefficiencies
  with witch,'' in \emph{Proceedings of the Twenty-Third International
  Conference on Architectural Support for Programming Languages and Operating
  Systems}, ser. ASPLOS '18.\hskip 1em plus 0.5em minus 0.4em\relax New York,
  NY, USA: ACM, 2018, pp. 332--347.

\bibitem{feather}
\BIBentryALTinterwordspacing
M.~Chabbi, S.~Wen, and X.~Liu, ``Featherlight on-the-fly false-sharing
  detection,'' in \emph{Proceedings of the 23rd ACM SIGPLAN Symposium on
  Principles and Practice of Parallel Programming}, ser. PPoPP '18.\hskip 1em
  plus 0.5em minus 0.4em\relax New York, NY, USA: Association for Computing
  Machinery, 2018, p. 152–167. [Online]. Available:
  \url{https://doi.org/10.1145/3178487.3178499}
\BIBentrySTDinterwordspacing

\bibitem{chromeprofiler}
``{Chrome Performance Profiler},''
  \url{https://developer.chrome.com/docs/devtools/}.

\bibitem{pyinstrument}
``{pyinstrument},'' \url{https://pyinstrument.readthedocs.io/en/latest/}, 2022.

\bibitem{python-wasm}
``{Wasmer Python},'' \url{https://github.com/wasmerio/wasmer-python}.

\bibitem{flamegraph}
G.~Brendan, ``{Flame Graphs},''
  \url{https://www.brendangregg.com/flamegraphs.html}.

\bibitem{spark}
\BIBentryALTinterwordspacing
M.~Zaharia, R.~S. Xin, P.~Wendell, T.~Das, M.~Armbrust, A.~Dave, X.~Meng,
  J.~Rosen, S.~Venkataraman, M.~J. Franklin, A.~Ghodsi, J.~Gonzalez,
  S.~Shenker, and I.~Stoica, ``Apache spark: A unified engine for big data
  processing,'' \emph{Commun. ACM}, vol.~59, no.~11, p. 56–65, oct 2016.
  [Online]. Available: \url{https://doi.org/10.1145/2934664}
\BIBentrySTDinterwordspacing

\bibitem{spark-bench}
``{Spark-Bench},'' \url{https://codait.github.io/spark-bench/}.

\bibitem{rdd}
``{Spark RDD},'' \url{https://data-flair.training/blogs/spark-rdd-tutorial/}.

\bibitem{dataset}
``{Spark Dataset},''
  \url{https://data-flair.training/blogs/apache-spark-dataset-tutorial/}.

\bibitem{intellijsdk}
``{IntelliJ Platform SDK},''
  \url{https://plugins.jetbrains.com/docs/intellij/welcome.html}.

\bibitem{rpcx}
``{A popular RPC test suite},'' \url{https://github.com/rpcxio/rpcx-benchmark}.

\bibitem{grpc}
``{gRPC},'' \url{https://grpc.io}.

\bibitem{memleaks}
``{Finding and fixing memory leaks in Go},''
  \url{https://dev.to/googlecloud/finding-and-fixing-memory-leaks-in-go-1k1h}.

\bibitem{LULESH}
I.~Karlin, A.~Bhatele, J.~Keasler, B.~L. Chamberlain, J.~Cohen, Z.~DeVito,
  R.~Haque, D.~Laney, E.~Luke, F.~Wang, D.~Richards, M.~Schulz, and C.~Still,
  ``Exploring traditional and emerging parallel programming models using a
  proxy application,'' in \emph{27th IEEE International Parallel \& Distributed
  Processing Symposium (IEEE IPDPS 2013)}, Boston, USA, May 2013.

\bibitem{tcmalloc}
``{TCMalloc},'' \url{https://github.com/google/tcmalloc}.

\bibitem{easyview-ae}
``{EasyView Artifact},'' \url{https://doi.org/10.5281/zenodo.10305415}.

\bibitem{vscode-remote-ex}
``{Remote Development},''
  \url{https://marketplace.visualstudio.com/items?itemName=ms-vscode-remote.vscode-remote-extensionpack}.

\end{thebibliography}
\bibliographystyle{IEEEtran}
\end{document}